\documentclass{Interspeech2024}
\usepackage{hyperref}
\usepackage{graphicx}

\interspeechcameraready

\title{Text-aware and Context-aware Expressive Audiobook Speech Synthesis}

\name[affiliation={1}]{Dake}{Guo}
\name[affiliation={1}]{Xinfa}{Zhu}
\name[affiliation={2}]{Liumeng}{Xue}
\name[affiliation={1}]{Yongmao}{Zhang}
\name[affiliation={1}]{Wenjie}{Tian}
\name[affiliation={1,*}]{Lei}{Xie}

\address{
  $^1$Audio, Speech and Language Processing Group (ASLP@NPU), School of Computer Science, \\ Northwestern Polytechnical University, Xi'an, China\\
  $^2$School of Data Science, The Chinese University of Hong Kong,\\ Shenzhen (CUHK-Shenzhen), China }
\email{guodake@mail.nwpu.edu.cn, lxie@nwpu.edu.cn\thanks{* Corresponding author.}}

\begin{document}

\maketitle

\begin{abstract}
    Recent advances in text-to-speech have significantly improved the expressiveness of synthetic speech. 
    However, a major challenge remains in generating speech that captures the diverse styles exhibited by professional narrators in audiobooks without relying on manually labeled data or reference speech.
    To address this problem, we propose a text-aware and context-aware (TACA) style modeling approach for expressive audiobook speech synthesis. We first establish a text-aware style space to cover diverse styles via contrastive learning with the supervision of the speech style. 
    Meanwhile, we adopt a context encoder to incorporate cross-sentence information and the style embedding obtained from text. Finally, we introduce the context encoder to two typical TTS models, VITS-based TTS and language model-based TTS. 
    Experimental results demonstrate that our proposed approach can effectively capture diverse styles and coherent prosody, and consequently improves naturalness and expressiveness in audiobook speech synthesis~\footnote{Speech samples: \href{https://dukguo.github.io/TACA-TTS/}{https://dukguo.github.io/TACA-TTS/}}.
\end{abstract}

\section{Introduction}

With the development of sequence-to-sequence (seq2seq) neural approaches~\cite{DBLP:conf/interspeech/WangSSWWJYXCBLA17,DBLP:conf/nips/RenRTQZZL19,DBLP:conf/icml/KimKS21}, current text-to-speech (TTS) systems can generate human-like natural speech, with wide applications in voice assistant, navigation system, and audiobook production. Particularly, audiobook speech synthesis aims to synthesize expressive long-form speech from literary books, achieving efficient and cost-saving automated audio-content production. However, audiobook synthesis is more challenging due to the \textit{rich speaking styles} performed by professional narrators, \textit{context-aware expressiveness}, and \textit{long-form prosody coherence}~\cite{10389629}.


To synthesize speech with various styles for audiobooks, a straightforward method is to train a TTS model with large-scale style-labeled datasets~\cite{choi2019multi,li2021controllable,liu2021expressive}. However, this method heavily depends on the quality and precision of the pre-defined labels. To mitigate reliance on annotated data, some works have tried to capture a global style embedding from a given style reference speech rendering a desired style~\cite{DBLP:conf/icml/WangSZRBSXJRS18,bian2019multi,min2021meta} and then generate speech with the corresponding style. Furthermore, CLAM~\cite{DBLP:conf/interspeech/MengL0LSXSZM22} proposes to select multiple references focusing on the style-related information in the text. Other works have focused on decoupling a style representation from speech, achieving style controllable speech synthesis~\cite{10095776,fu2021bi,song2023multi}. However, the styles derived from either labeled tags or reference speech are still limited, making it difficult to cover the wide range of rich styles in audiobooks. Additionally, it is also impractical to manually assign a specific style tag or style reference for each sentence in an audiobook. 

A more practical method for style extraction for speech synthesis is to predict the style embedding from text because there are only text can be accessible during inference. To achieve such text-adaptive style learning for speech synthesis, Text-Predicted Global Style Token (TP-GST) extends the capabilities of GST\cite{DBLP:conf/icml/WangSZRBSXJRS18,DBLP:conf/slt/StantonWS18} to generate style embeddings or style token weights based on textual input solely. CLAPSpeech~\cite{ye-etal-2023-clapspeech} utilizes contrastive learning to learn the prosody variance of the same text under different contexts. However, it focuses on local prosody, such as word-level or even phoneme-level prosody. Even though TP-GST produces global style embedding, it relies on a limited number of style tokens, which is not enough for diverse style variance in audiobooks with abundant textual content.


Moreover, the style and prosody in each sentence are affected by its neighboring sentences in the case of long text in the book, and thus many methods improve the prosody consistency of the synthetic speech by learning contextual information~\cite{zhang2022maskedspeech}. Some of them use context-aware text embeddings of continuous multiple sentences to improve the prosody of speech~\cite{DBLP:conf/icassp/XuSZZHZ21,10129796}. Other works leverage acoustic contexts to improve the prosody coherence and expressiveness of the current sentence~\cite{DBLP:journals/corr/abs-2012-03763}. Moreover, some studies utilize both textual and acoustic contextual knowledge to generate expressive long-form speech~\cite{10095866,10096247}. Besides, given that the expressiveness of human speech is perceived as a composition of hierarchical factors rather than a singular scale, several studies leverage the hierarchical structure of discourse to further enrich the prosody variations in long-form synthetic speech~\cite{10389629,DBLP:journals/taslp/XueSZX22,DBLP:conf/icassp/LeiZCWKM22,DBLP:conf/interspeech/LeiZCH0KM22}.

This paper proposes a text-aware and context-aware (TACA) style modeling approach for expressive audiobook speech synthesis. To address the issues of limited styles and achieve more practical text-adaptive style prediction, we propose a text-aware style modeling method to learn a style space encompassing a wide range of styles. Specifically, we employ a semi-supervised contrastive learning strategy to learn style directly from the text using a large number of datasets. This process is assisted by the style learned from speech, as it allows for the capture of more detailed style variations in speech. To further integrate context, a context encoder is crafted to extract cross-sentence information from adjacent sentences and to embed the learned style directly from the text, assisting TTS in generating speech with context-aware expressiveness and long-form coherent prosody. Since the context encoder is a plug-and-play, universal module, it can be readily integrated into most TTS frameworks. In this study, we build text-aware and context-aware (TACA) TTS models based on two different TTS frameworks, i.e., a classical VITS-based TTS and a popular language model-based TTS. Experimental results demonstrate that the text-aware style modeling method effectively captures the diverse and detailed style from text and the context encoder helps to learn coherent prosody. 

\section{Methodology}
\begin{figure}[ht]
  \centering
  \includegraphics[width=\linewidth]{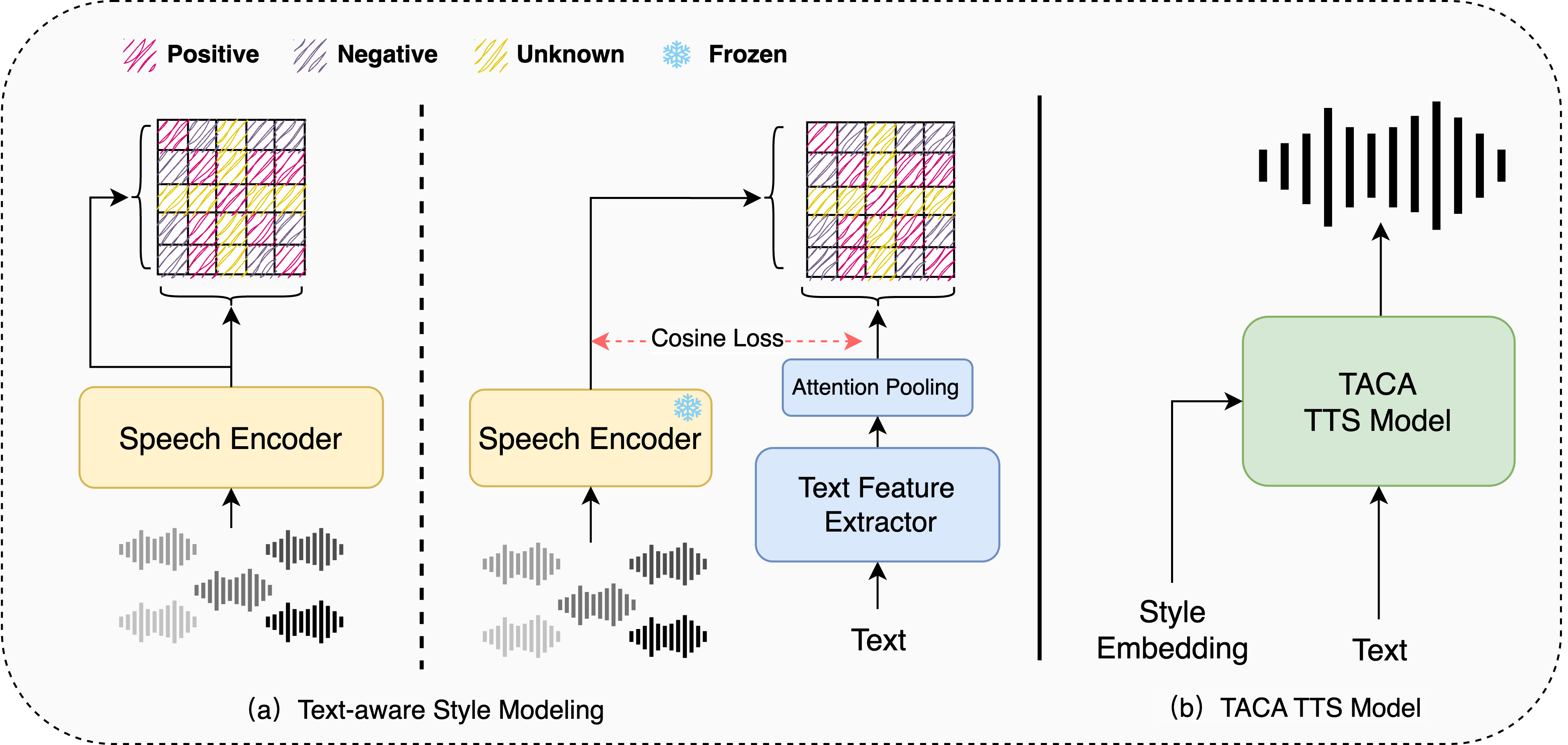}
  \caption{Overview of our proposed method}
  \label{fig:over}
  \vspace{-10pt}
\end{figure}

The overall architecture of our proposed method is shown in Figure~\ref{fig:over}. We first construct an extensive style space from speech and then help to establish text-aware style space using cross-modal supervision. Then we build text-aware and context-aware (TACA) TTS models that incorporate the context encoder with the style embedding obtained from the text, to introduce contextual information and diverse styles into TTS for improved audiobook speech synthesis in expressiveness and naturalness.

\subsection{Text-aware Style Modeling}

Given speech-text pairs $P_1, P_2, ... P_n$, where speech-text pair $P_i$ contains speech $S_i$ and text $T_i$, we aim to build a style space from text to speech. Considering the style expressions mainly expressed through speech, we first establish a speech style space and then help the text style space learning under cross-modal supervision. To capture rich style expressions in speech $S_1, S_2, ... S_n$, we utilize the speech encoder from SRL~\cite{zhu2023multi} to extract style representations. Specifically, this speech encoder leverages HuBERT~\cite{hsu2021hubert} to handle various speech inputs and uses contrastive learning to capture style representation in a semi-supervision fashion. Compared to classifier-guided representation learning models that categorize styles into predefined classes, the style representations learned from stylistic differences between speech samples offer greater diversity, contributing to a more detailed and comprehensive style space. 


Under the guidance of the established speech style space, we build the text-aware style space on a larger amount of datasets. Inspired by CLIP~\cite{radford2021learning}, we leverage contrastive learning to assist text-style space learning with the learned speech-style space. The key to contrastive learning lies in constructing positive and negative samples. We construct these samples for text-style learning based on the similarity between the learned speech representations. If the similarity between $S_i$ and $S_j$ is less than $\alpha$, the pairs $P_i$ and $P_j$ are denoted as negative samples. Conversely, if the similarity between $S_i$ and $S_j$ is more than $\beta$, the pairs $P_i$ and $P_j$ are denoted as positive samples. 
Any pairs with similarity scores ranging between $\alpha$ and $\beta$ are termed unknown samples, reflecting ambiguity in their positive or negative correlation. As shown in Figure~\ref{fig:over} (b), we employ a powerful pre-trained text feature extractor, such as the T5 model, and attention pooling to obtain global style representations from $T_1, T_2, ... T_n$. Then we fine-tune the text feature extractor but freeze the speech encoder to learn the text-aware style via calculating the contrastive loss of $P_1, P_2, ... P_n$ in a semi-supervised manner. Furthermore, we use cosine similarity loss on the style embeddings of the two modalities to accelerate convergence. Such a style modeling method not only learns rich and diverse styles via contrastive learning and semi-supervision modes but also makes the style text adaptive, facilitating the varied and expressive expression in audiobook speech synthesis where only text is available.

\subsection{Context Enocder}
\begin{figure}[h]
  \centering
\vspace{-10pt}\includegraphics[width=0.6\linewidth]{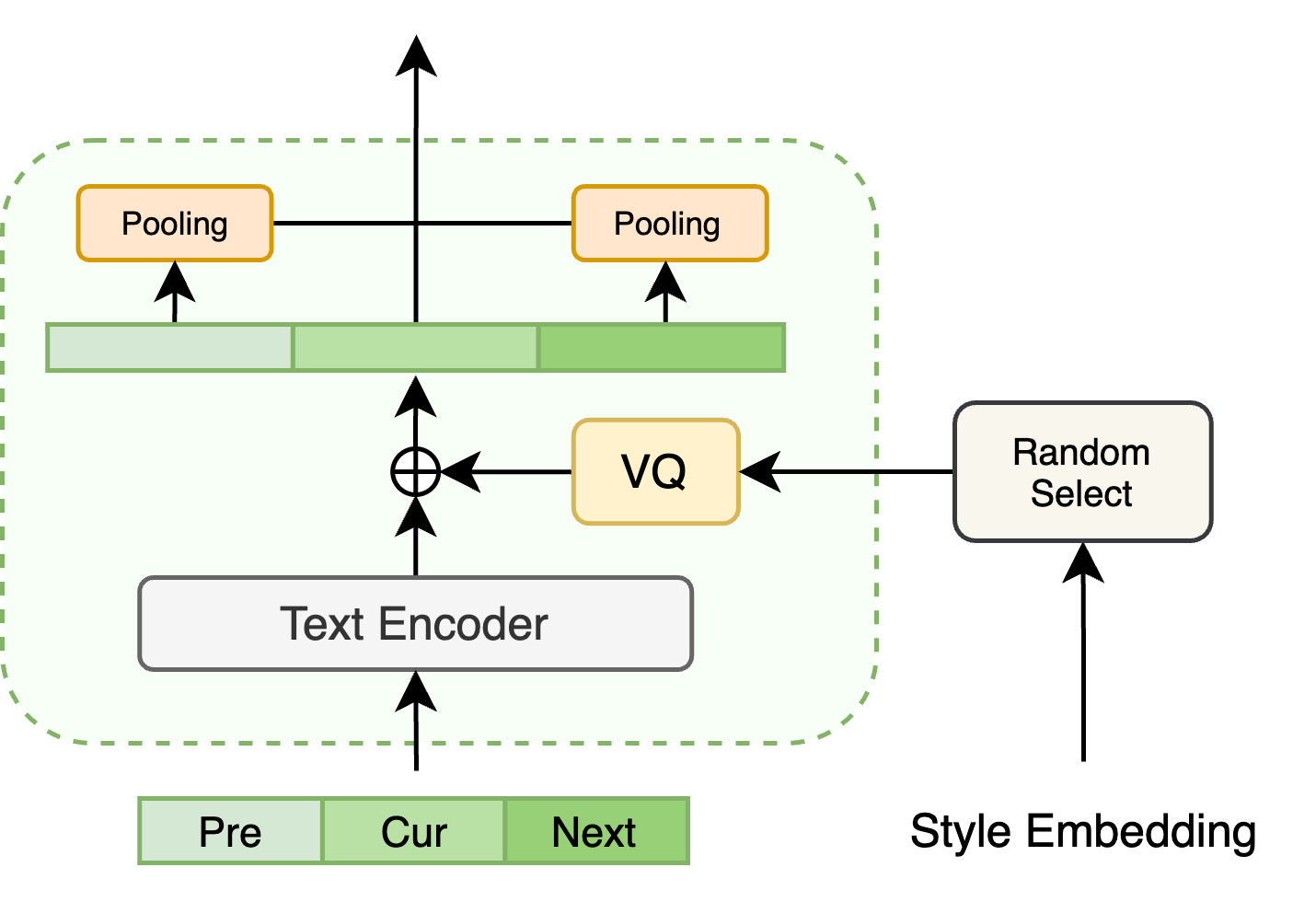}
  \caption{The architecture of context encoder}
  \label{fig:ce}
\vspace{-10pt}
  
\end{figure}

We design a context encoder to integrate contextual information with style representations, as shown in Fig~\ref{fig:ce}. The prosody of a given sentence can be influenced by its position within a discourse and the relationships with adjacent sentences. Considering this, we broaden the receptive field of the text input in TTS models from one sentence to a couple of sentences. It explicitly incorporates context from neighboring sentences. Furthermore, we introduce the style embedding obtained from the text-aware style space to the context encoder to encourage the TTS model to generate expressive speech. It is worth noting that we apply Vector Quantization (VQ) to the style embedding. It enables the model to effectively learn and capture the commonalities in style expression across a wide range of samples, generating more stable expressive speech.


To make full use of the available dataset including the dataset with and without contextual annotation which is the order of the utterance, we adopt a basic pre-training and \textit{context-aware fine-tuning} training strategy to use the context encoder. Specifically, the pre-training stage is to train the model without contextual information, and the fine-tuning stage is to train the model with context information. During the fine-tuning process, style embeddings are randomly sourced from either speech or text to bolster the model's generalization. 


\subsection{Text-aware and Context-aware TTS}
We apply the context encoder with text-aware style and contextual information to two typical types of TTS models, i.e., the VITS-based model and the language model-based model, to achieve expressive audiobook speech synthesis.

\subsubsection{VITS-based TTS}
\begin{figure}[ht]
  \centering
  \includegraphics[width=0.8\linewidth]{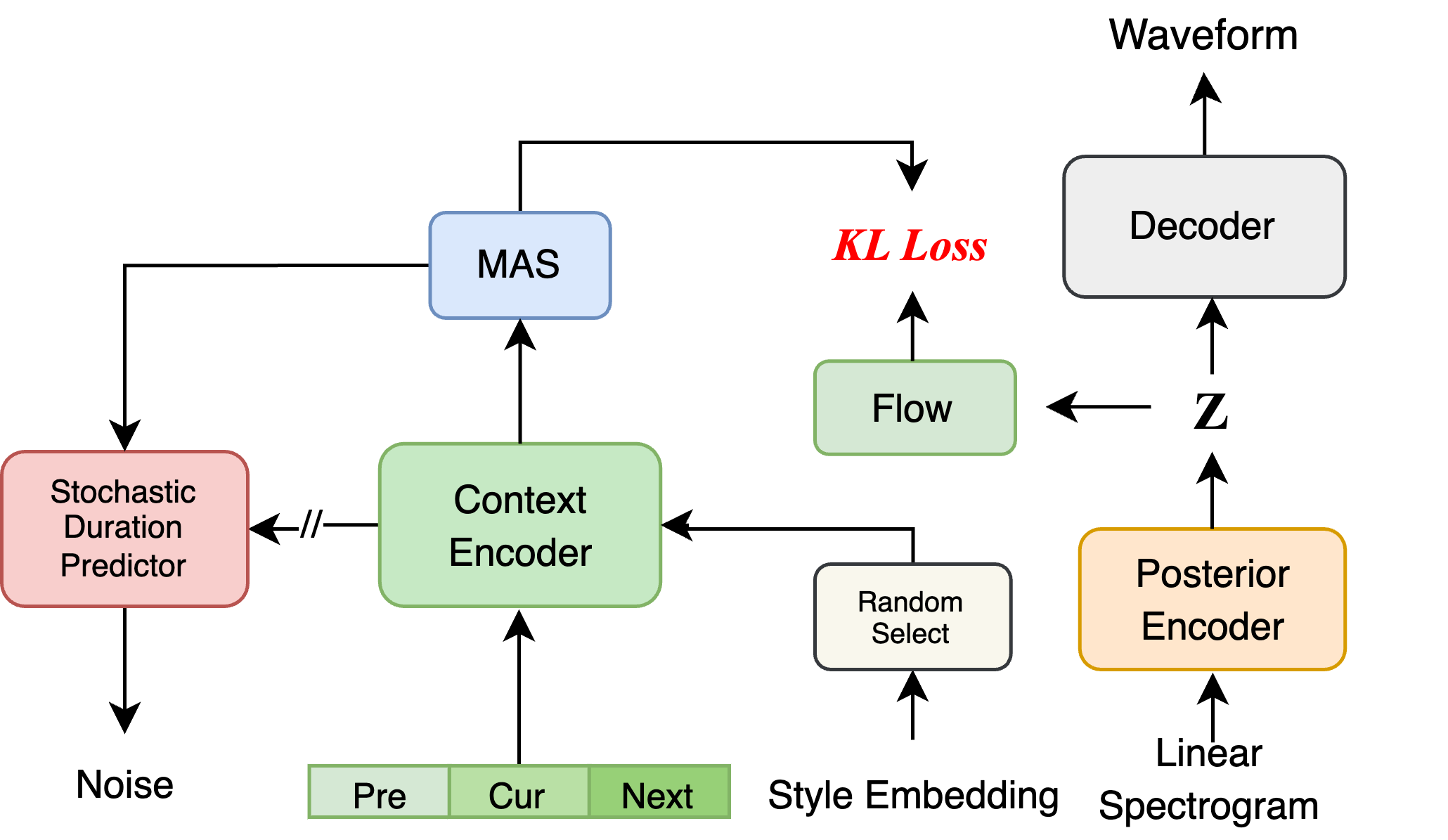}
  \caption{Text-aware and context-aware VITS-based TTS}
  \label{fig:vits}
\vspace{-10pt}
  
\end{figure}

We employ BERT-VITS 2~\footnote{\href{https://github.com/fishaudio/Bert-VITS2}{https://github.com/fishaudio/Bert-VITS2}} as one of the TTS models, acknowledged for its exceptional expressiveness in speech synthesis. Bert-VITS 2, building upon the VITS 2~\cite{kong23_interspeech} architecture, incorporates BERT embeddings, enriching the synthetic speech with semantic information and significantly improving its prosody. We replaced the original text encoder with the context encoder, achieving a larger receptive field of context to better model long-form speech, as shown in Figure~\ref{fig:vits}. We apply basic pre-training and \textit{context-aware fine-tuning} training strategy to obtain a Base-VITS and a TACA-VITS, where the TACA-VITS is augmented with rich style and contextual information for more expressive speech synthesis.


\subsubsection{Language model-based TTS}

\begin{figure}[h]
  \centering
  \includegraphics[width=0.8\linewidth]{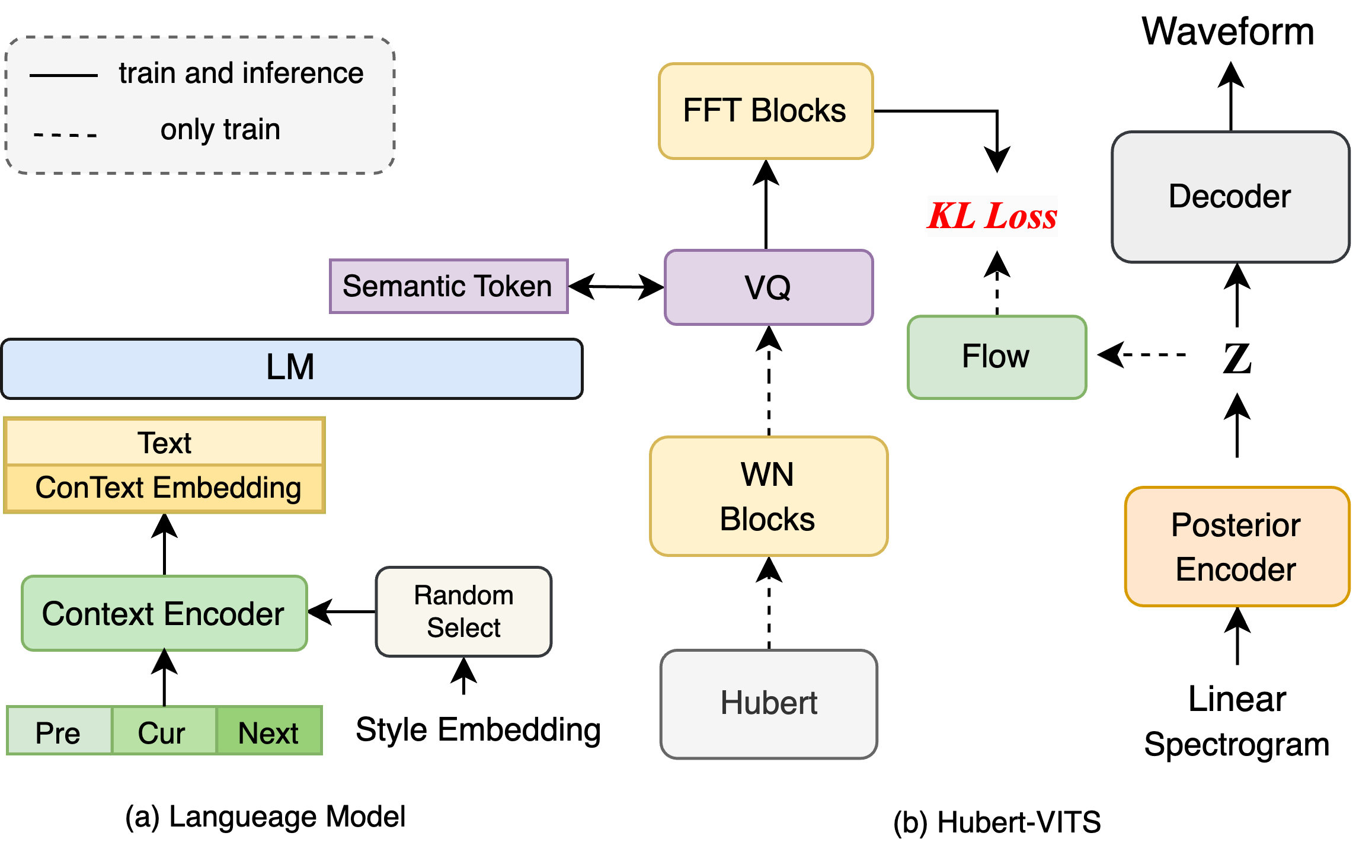}
  \caption{Text-aware and context-aware LM-based TTS}
  \label{fig:lm}
\vspace{-10pt}
  
\end{figure}

Leveraging language models for speech synthesis based on discrete tokens has evolved into a new research paradigm. We build a language model-based TTS as another TTS model, as shown in Figure~\ref{fig:lm}, which is similar to AR-VITS~\footnote{\href{https://github.com/innnky/ar-vits}{https://github.com/innnky/ar-vits}}. As shown in the right part of Figure~\ref{fig:lm}, it begins to extract audio features using the Hubert model with WN Blocks~\cite{DBLP:conf/icml/KimKS21}, followed by the application of VQ for semantic tokens, and the tokens are transformed into waveforms using VITS, which is called Hubert-VITS. In the Hubert-VITS model, the input consists of frame-level tokens, thus eliminating the necessity for duration modeling to stretch the input to a frame-level representation. The left part of Figure~\ref{fig:lm} is an autoregressive (AR) language model, which aims to predict the semantic tokens from the text.
In this work, we first train the Hubert-VITS model using high-quality audio samples, aiming to enhance its waveform reconstruction capabilities. Subsequently, we pre-train a transform-based language model to predict semantic tokens on a large dataset because language models perform better on big datasets. After that, we apply \textit{context-aware fine-tuning} (TACA-LM) to the pre-trained language model (Base-LM) to encourage the language model with a larger receptive field of context and more detailed style information, improving the expressiveness of the synthesized speech.

\section{Experimental Setup}

\subsection{Datasets and Preprocessing}
Our validation methodology for the framework involves the application of audiobooks and an expressive speech corpus. There are three datasets involved in experiments.
\begin{itemize}
\item \textbf{20H-Audiobook-HQ:} A Mandarin audiobook dataset with utterance order annotated contains about 20 hours of speech with the corresponding text.
\item \textbf{100H-Multi-Style:} A multi-style Mandarin dataset contains about 100 hours of speech with the corresponding text, of which 66 hours are labeled with the style tags. 
\item \textbf{6kH-Audiobook:} An audiobook dataset from the Internet contains about 6000 hours of speech without text. We transcribe the audio to text using Paraformer~\cite{gao2022paraformer}.
\end{itemize}
We select 5 chapters, about 100 audio clips from 20H-Audiobook-HQ as the test set.

\subsection{Training Setup and Model Details}




During text-aware style modeling, we use the Chinese-Hubert-Large~\footnote{\href{https://github.com/TencentGameMate/chinese_speech_pretrain}{https://github.com/TencentGameMate/chinese\_speech\_pretrain}} to extract features from speech with layer 6. We use the same configuration as SRL in Speech Encoder. For text feature extractor, we use a pre-trained Chinese T5, Randeng-T5-784M~\footnote{\href{https://huggingface.co/IDEA-CCNL/Randeng-T5-784M}{https://huggingface.co/IDEA-CCNL/Randeng-T5-784M}}. The $\alpha$ and $\beta$ in text-aware style modeling are set to 0.60 and 0.95, respectively. The dimension of style embedding is set to 384. All embeddings from either Speech Encoder or T5  undergo L2 normalization before calculating loss. We train the Speech Encoder on 100H-Multi-Style  with a batch size of 96 and T5 on 6kH-Audiobook with a batch of 64.

For TACA-VITS, the model configuration is consistent with the original repository, which is the version of 2.0. The Context Encoder takes contextual phonemes and aligned Bert embeddings as input. The vector quantization of style in Context Encoder uses a codebook with 64*32.
We train the base model using 100H-Multi-Style with a batch size of 48. When \textit{context-aware fine-tuning}, we train the model in the same setup.

For TACA-LM, we extract semantic tokens with vector quantization from layer 9 of the Chinese-Hubert-Base. The codebook size and dim is set to 1024 and 128, respectively. The configuration of VITS is the same as the original paper. The Hubert-VITS is trained with a batch size of 48 using 300 hours of high-quality speech selected from 6kH-Audiobook. We employ nanoGPT~\footnote{\href{https://github.com/karpathy/nanoGPT}{https://github.com/karpathy/nanoGPT}} as the backbone of the language model, which contains 12 transformer layers with 12 heads. The input sequence applies only one position embedding, and its embedding dimension is set to 768. Different from traditional AM, we employ text as input rather than phonemes. The language model is trained on 6kH-Audiobook with a batch size of 24. Also, 10 times gradient accumulation is used during training. The context encoder is trained on the same configuration as TACA-VITS. 

\section{Experimental Results}
To ascertain the efficacy of the style embedding learned via cross-modal supervision, we first conduct a detailed analysis of the style space from the text style encoder. Subsequently, we further investigate the effectiveness of the two TTS models on audiobook synthetic speech in expressiveness and naturalness compared with vanilla TTS models without context encoder.




\begin{figure}[h]
  \centering
  \includegraphics[width=0.68\linewidth]{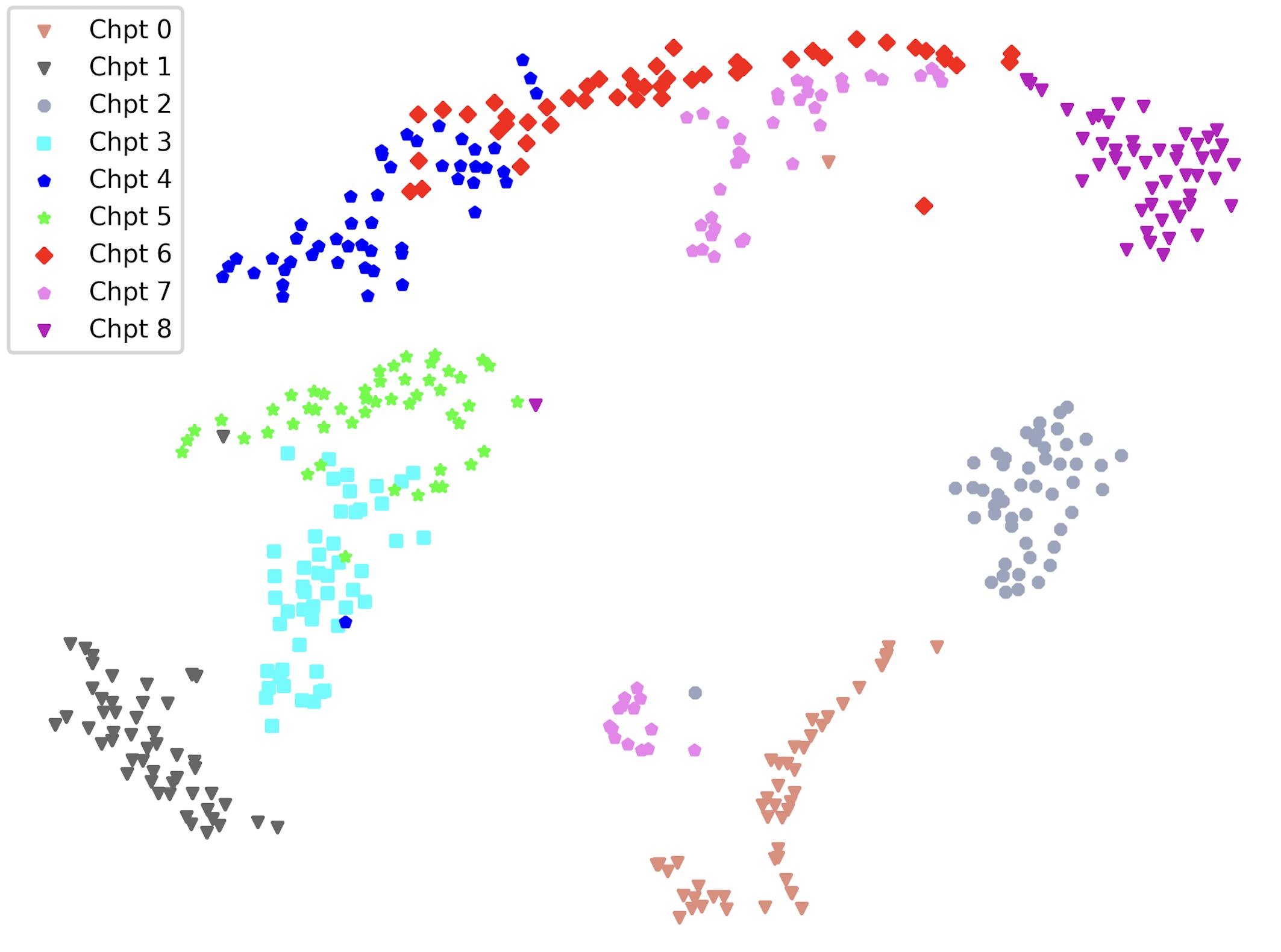}
  \caption{t-SNE visualization of the style representation extracted from the text in 9 different chapters.}
  \label{fig:tsne}
  \vspace{-15pt}
\end{figure}



\subsection{Style Space Analysis}
The speaking style is mainly expressed through speech. In the text-speech paired dataset, the style predicted from the text should be aligned with that of the corresponding speech. To verify whether the style space built by our cross-modal style modeling method is shared between text and speech, we calculate the cosine similarity between the style embeddings extracted from speech and text on 1000 sentences. The higher cosine similarity of 0.93 indicates that our method is effective in aligning the text to the speaking style in the speech. Moreover, the T5 encoder used for text feature extraction is more powerful than BERT which obtains the cosine similarity of 0.82, benefitting to the style embedding predicted from text and then resulting in expressive generated speech. 

We further visualize the style representations predicted from the text of 9 chapters, including 450 sentences in total. The t-SNE visualization result is presented in Figure~\ref{fig:tsne}, in which points with the same color are derived from the same chapters. From the visualization, we can see that (1) The style embedding of text from the same chapter tends to cluster a group, indicating that our method captures a relatively consistent style within a chapter. It helps to produce coherent prosody in audiobook speech synthesis. (2) The points in a cluster group are relatively dispersed, which means the nuance in style among sentences within a chapter can be learned deliberately. (3) The points in the whole style space are diversely distributed, showing that the style space covers diverse styles. It benefits audiobook speech synthesis with extensive expressiveness.



\subsection{Objective Evaluation}
We evaluate the performance of TTS models on the reserved test set with character error rate (CER) and mel-cepstral distortion (MCD). We use Paraformer~\cite{gao2022paraformer} to evaluate CER and employ dynamic time warping (DTW) to align the predicted mel-spectrogram with the ground truth to calculate the MCD. The objective results are shown in Table \ref{obj}, where VITS and LM refer to the vanilla methods, and both of them are fine-tuned on 6kH-Audiobook for a fair comparison. 
As we can see, LM is somewhat lacking in quality and stability compared to VITS reflected in higher MCD and CER. However, after context-aware fine-tuning of VITS (TACA-VITS), the overall changes in MCD and WER are not significant, indicating that our training strategy doesn't degrade model performance. For LM, applying the same fine-tuning strategy (TACA-LM) results in reduced MCD and CER, which implies that for LMs with stronger modeling capabilities, integrating conditions associated with semantics and style can slightly improve speech quality and stability. Compared to the two models of the VITS,  the LM-based methods demonstrate higher CER, chiefly due to the inherent problems of autoregressive models, including missed readings and repetition.


\begin{table}[h]
\vspace{-5pt}

\centering
\caption{ Evaluations results for different TTS models.}
\label{obj}
\resizebox{0.9\linewidth}{!}{
\begin{tabular}{r|cc|cc}
\toprule
\textbf{TTS Model} & MCD $\downarrow$  &CER(\%) $\downarrow$ & NMOS $\uparrow$  & EMOS $\uparrow$\\ \midrule
VITS            &    4.23  &             \textbf{5.80}                & 3.84$\pm$0.110  &  3.61$\pm$0.143             \\ 
TACA-VITS         & \textbf{4.21}     &   5.99                          &   \textbf{3.90$\pm$0.098}  & \textbf{3.93$\pm$0.105  }                  \\ \hline
LM           &  8.92   & 13.9   & 2.91$\pm$0.087   & 3.80$\pm$0.112 \\
TACA-LM              & \textbf{8.15 }     & \textbf{13.1} &\textbf{3.22$\pm$0.099}    &\textbf{4.05$\pm$0.104}  \\
\bottomrule
\end{tabular}
}
\vspace{-15pt}
\end{table}

\subsection{Subjective Evaluation}
We evaluate synthetic audiobook speech in terms of naturalness and expressiveness through Mean Opinion Score (MOS) tests, denoted as EMOS and NMOS, respectively. NMOS assesses perceptual quality, including clarity and intelligibility, while EMOS focuses on prosodic expression, ignoring quality issues.  To align closely with the audiobook scenario, we generate five audio clips with lengths ranging from 1 to 2 minutes constructed by combining multiple sentences for each MOS test. In each MOS test, a group of 20 native Chinese Mandarin listeners is asked to listen to synthetic speech and rate on a scale from 1 to 5 with a 0.5-point interval.

The MOS results presented in Table~\ref{obj} show that our proposed method achieves improvements in naturalness and expressiveness on both two TTS models. 
The observed gains in NMOS and EMOS metrics underscore the effectiveness of our proposed method, particularly highlighting the pivotal role of the context encoder. 
After incorporating context information, TACA-VITS shows a slight improvement over VITS in naturalness, and TACA-LM also exhibits an improvement in naturalness. It indicates that context-aware knowledge helps to coherent prosody and results in more natural speech.
Additionally, TACA-VITS demonstrates a significant improvement in expressiveness, and TACA-LM also sees a notable improvement in expressiveness. The contextual information and text-adaptive style guidance bring context-aware expressiveness and rich styles, producing more expressive audiobook speech.
It is noted that the VITS models demonstrate superior performance over LM models in naturalness, which can be attributed to the inherent backdraws of autoregressive inference in LM, such as repetition, omission, and unclear pronunciation. However, LM shows great potential in expressiveness owing to its excellent capabilities of semantic comprehension and context-aware style expression.



\section{Conclusions}
This paper proposes a novel method for expressive audiobook speech synthesis through text-aware and context-aware style modeling.  We first establish a text-aware style space to cover diverse styles with the supervision of speech style space. Furthermore, we design a context encoder to incorporate contextual information in cross-sentence and embed style embeddings derived from text. Since text-aware style modeling and context-aware information learning methods are designed universally, they can be integrated into most TTS frameworks. We build VITS-based and LM-based TTS systems based on the two modules. Experimental results show that our method effectively captures diverse styles and coherent prosody, improving naturalness and expressiveness in audiobook speech synthesis.

\bibliographystyle{IEEEtran}
\bibliography{mybib}

\end{document}